# Perceptions of the Schrodinger Equation

Spyros Efthimiades

*Fordham University, Department of Natural Sciences,
New York, NY 10023, U.S.A.* sefthimiades@fordham.edu

July 5, 2013

**Abstract**

The Schrodinger equation has been considered to be a postulate of quantum physics, but it is also perceived and derived heuristically as the quantum equivalent of the classical energy relation. We indicate that the Schrodinger equation cannot be a physical postulate, and we show that the heuristic derivations wrongly consider that the particle kinetic energy is proportional to the square of the momentum operator acting on the wavefunction. The correct term is proportional to the square of the momentum. Analyzing particle interactions, we realize that particles have multiple virtual motions and that each motion is accompanied by a wave that has constant amplitude. Accordingly, we define the wavefunction as the superposition of the virtual waves of the particle. As a result, quantum mechanics becomes a local and global theory, while in the traditional formulation it is a local theory. In several interaction settings we can tell what particle motions arise and explain the outcomes in simple and tangible terms. Most importantly, the mathematical foundation of quantum mechanics becomes clear and justified, and we derive the Schrodinger, Dirac, etc. equations as the conditions the wavefunction must satisfy at each space-time point in order to fulfill the respective total energy equation.

PACS numbers: 03.65.−w.

Key words: Schrodinger Equation, principles of Quantum Mechanics

# 1. INTRODUCTION

The Schrodinger equation is a space-time differential equation that, within its limits of accuracy and applicability, yields the wavefunctions of particles interacting with potentials. The Schrodinger equation is considered to be a postulate of quantum physics that "cannot be derived from simpler principles" [1], but it is also assumed to be the quantum equivalent of the classical energy equation from which it is derived heuristically. We will show that these perceptions are flawed and that they wrongly associate the particle kinetic energy at a point with $-\hbar^2(2m)^{-1}\psi^*\nabla^2\psi$. The correct term is $+\hbar^2(2m)^{-1}\nabla\psi^*\cdot\nabla\psi$.

In section 2, we present the reasons why the Schrodinger equation cannot be a physical postulate. In section 3, we show that the heuristic derivations are inappropriate. In section 4, we present the correct local kinetic energy term and show explicitly that the assumed form yields incorrect values. In section 5, we discuss the deficiencies of the traditional quantum theory.

In section 6, we analyze particle interactions and conclude that particles have multiple simultaneous virtual motions, and that each motion is accompanied by a wave that has constant amplitude. Accordingly, we define the particle wavefunction as the superposition of the virtual waves of the particle. Based on this definition, quantum theory becomes clear and justified and, in section 7, we describe particle interactions in qualitative terms. In section 8, we derive the Schrodinger, Dirac, etc. equations as the local conditions the wavefunction must satisfy in order to fulfill the corresponding total energy equation.

In section 9, we discuss the local character of traditional quantum mechanics and the local and global quantum theory that emerges from an proper representation of particle interactions, that shows the physical content of the wavefunction equations and leads to their derivation.

In order to have the simplest possible mathematical expressions, we will refer mostly to the one-dimensional, time-independent energy and Schrodinger equations:

$$E(x) = \frac{p^2(x)}{2m} + V(x) \qquad\qquad E\psi = \frac{-\hbar^2}{2m}\frac{\partial^2\psi}{\partial x^2} + V(x)\psi \qquad (1)$$

# 2. AXIOMATIC APPROACH

Textbooks introduce the Schrodinger equation as a postulate of quantum physics that "plays a role analogous to that of Newton's second law of motion in classical mechanics" [1]. This analogy is incorrect. There are several reasons why the Schrodinger equation cannot be a physical principle. For example:
*i.* The Schrodinger equation has the form of the non-relativistic energy equation; therefore, it must be a derivative condition related to the energy equation.
*ii.* Due to its limitations of applicability and accuracy, the Schrodinger equation does not qualify as a physical principle.
*iii.* The Schrodinger equation is put in the following form in which it is claimed to be the principle that provides the time evolution of the particle wavefunction:

$$i\hbar\frac{\partial\psi}{\partial t} = \hat{H}\psi \qquad (2)$$

where $\hat{H}$ is the "Hamiltonian operator". However, the Hamiltonian is defined to be the kinetic plus potential energies, which is not the case for the Schrodinger equation since, as we will see, the term that contains the second space derivative does not correspond to the kinetic energy.

Moreover, we should bear in mind that the fundamental physical quantity is the particle wavefunction, not the Schrodinger, Pauli, Dirac, etc. equations from which it is obtained in certain cases. However, by defining the Schrodinger equation as a postulate, the physical origin of the wavefunction becomes undetermined since it is the solution of a postulated equation.

## 3. HEURISTIC DERIVATIONS

In most textbooks, the Schrodinger equation is derived heuristically as the quantum equivalent of the local classical energy equation. In fact, there are two main heuristic approaches.

**A. Applying the "Kinetic Energy Operator":** The wavefunction equations are derived, heuristically, by multiplying the classical energy equations by the wavefunction $\psi$ and substituting the energy, momentum, etc. variables by the corresponding quantum operators $i\hbar\partial_t, -i\hbar\nabla, etc.$ The "kinetic energy operator" is $-\hbar^2(2m)^{-1}\nabla^2$ [2]. According to this recipe, $p^2(x)$ is substituted by $\hat{p}^2 = -\hbar^2 \partial_x^2$ in the energy equation in (1), and we get the Schrodinger equation also shown in (1). Multiplying this equation by $\psi^*$, we have

$$E|\psi|^2 = \frac{-\hbar^2}{2m}\psi^* \frac{\partial^2\psi}{\partial x^2} + V(x)|\psi|^2 \qquad (3)$$

It is generally assumed that the term containing the second space derivative is the kinetic energy of the particle at $x$, but we will show that this is not the case. Thus, the "kinetic energy operator" does not give the correct kinetic energy term. In conclusion, this heuristic derivation is based on a wrong premise.

**B. Extending the "Plane Wave Equation":** Many textbooks, i.e. [3], use the following method to derive the Schrodinger equation.

They start by representing a single plain wave of a free particle by

$$\psi_p = Ae^{\frac{i}{\hbar}(px-Et)} \qquad (4)$$

which can be the solution of the following wave equation

$$i\hbar\frac{\partial\psi_p}{\partial t} = \frac{-\hbar^2}{2m}\frac{\partial^2\psi_p}{\partial x^2} \quad \rightarrow \quad E\psi_p = \frac{-\hbar^2}{2m}\frac{\partial^2\psi_p}{\partial x^2} \quad where \quad E = \frac{p^2}{2m} \qquad (5)$$

Next, they claim that when a particle interacts with a potential $V(x)$, we can substitute $E$ with $[E - V(x)]$ in equation (5) and obtain the Schrodinger equation

$$[E - V(x)]\psi(x) = \frac{-\hbar^2}{2m}\frac{\partial^2\psi(x)}{\partial x^2} \qquad (6)$$

This substitution is inappropriate for a host of reasons. For example:
i. While the term $-\hbar^2(2m)^{-1}\psi^*\partial_x^2\psi$ does represent the kinetic energy of a single particle wave $\psi_p$, we will show that − in spite of its plausible form − it does not represent the particle kinetic energy when $\psi$ is a wavefunction.
ii. The analogy used to convert the single wave equation (5) into the wavefunction equation (6) is flawed because the initial wave equation does not have any predictive power; it just gives back what we have already put in. Extending a simple relation to obtain an equation, that cannot be justified, is not acceptable.

iii. The particle wave $\psi_p$ and the wavefunction $\psi$ are mathematically dissimilar functions because the former has constant intensity and the latter variable. So the two equations that describe $\psi_p$ and $\psi$ cannot be validly related.

In conclusion, the simplistic extension of the single plane wave equation to the Schrodinger wavefunction equation constitutes a physically and mathematically unjustified jump.

## 4. THE KINETIC ENERGY TERM

Equation (3) is an energy condition. However, while $E|\psi(x)|^2$ and $V(x)|\psi(x)|^2$ correspond to the particle energy and potential energy, respectively, the third term is not the kinetic energy of the particle.

Integrating equation (3) over space we get the total energy equation

$$E = \int E|\psi|^2 dx = \int \frac{-\hbar^2}{2m}\psi^* \frac{\partial^2 \psi}{\partial x^2} dx + \int V(x)|\psi|^2 dx$$
$$= \int \frac{+\hbar^2}{2m} \frac{\partial \psi^*}{\partial x} \frac{\partial \psi}{\partial x} dx + \int V(x)|\psi|^2 dx = KE + PE \tag{7}$$

where we have added the vanishing surface term that appears below.

$$\int \frac{+\hbar^2}{2m} \frac{\partial \psi^*}{\partial x} \frac{\partial \psi}{\partial x} dx = \int \frac{+\hbar^2}{2m} \frac{\partial \psi}{\partial x} d\psi^* = \frac{+\hbar^2}{2m} \psi^* \frac{\partial \psi}{\partial x}\bigg|_S + \int \frac{-\hbar^2}{2m}\psi^* \frac{\partial^2 \psi}{\partial x^2} dx = \int \frac{-\hbar^2}{2m}\psi^* \frac{\partial^2 \psi}{\partial x^2} dx \tag{8}$$

We point out that this surface term can be zero on surfaces enclosing small volumes if the intensity of $\psi$ is constant. This is the case for a single wave $\psi_p$, but not for a wavefunction. Thus, for $\psi_p$ we have

$$\frac{+\hbar^2}{2m} \frac{\partial \psi_p^*}{\partial x} \frac{\partial \psi_p}{\partial x} = \frac{-\hbar^2}{2m} \psi_p^* \frac{\partial^2 \psi_p}{\partial x^2} = \frac{+p^2}{2m}|\psi_p|^2 \tag{9}$$

The wavefunction $\psi(x)$ yields the probability that the particle may be at point $x$ but also the particle momentum $p$ and energy $E$ at $x$. In traditional quantum mechanics, the momentum of the particle at $x$ is obtained by applying the momentum operator on the wavefunction:

$$p(x) = \hat{p}\psi = -i\hbar \frac{\partial \psi}{\partial x} \tag{10}$$

The above expression is true for a single particle wave. In section 7, we will show that it is also valid when $\psi$ is a wavefunction.

The particle kinetic energy $KE(x)$ is proportional to $|\hat{p}\psi|^2$. However, the traditional quantum mechanics postulates that the square of the momentum is obtained by applying the square of the momentum operator on the wavefunction and so the corresponding "kinetic energy term" is $K(x)$, proportional to $|\hat{p}^2\psi|$. Specifically, we have

$$KE(x) = \frac{1}{2m}|\hat{p}\psi|^2 = \frac{1}{2m}\left(+i\hbar\frac{\partial\psi^*}{\partial x}\right)\left(-i\hbar\frac{\partial\psi}{\partial x}\right) = \frac{+\hbar^2}{2m}\frac{\partial\psi^*}{\partial x}\frac{\partial\psi}{\partial x}$$

$$K(x) = \frac{1}{2m}\psi^*\hat{p}^2\psi = \frac{1}{2m}\psi^*\left(-i\hbar\frac{\partial}{\partial x}\right)\left(-i\hbar\frac{\partial}{\partial x}\right)\psi = \frac{-\hbar^2}{2m}\psi^*\frac{\partial^2\psi}{\partial x^2}$$

(11)

To demonstrate the difference between $KE(r)$ and $K(r)$, we evaluate them for the electron ground eigenstate $\Psi_1(r)$ in the Hydrogen atom, for which we have:

$$\Psi_1(\mathbf{r}) = \frac{1}{\sqrt{\pi a^3}}e^{-\frac{r}{a}} \qquad a = \frac{\hbar^2}{me^2}$$

$$KE(r) = \frac{+\hbar^2}{2m}\nabla\Psi_1^*\cdot\nabla\Psi_1 = \frac{1}{\pi a^3}\left[\frac{e^2}{2a}\right]e^{-\frac{2r}{a}}$$

$$K(r) = \frac{-\hbar^2}{2m}\Psi_1^*\nabla^2\Psi_1 = \frac{1}{\pi a^3}\left[\frac{e^2}{r} - \frac{e^2}{2a}\right]e^{-\frac{2r}{a}}$$

(12)

$$PE(r) = V(r)|\Psi_1|^2 = \frac{1}{\pi a^3}\left[-\frac{e^2}{r}\right]e^{-\frac{2r}{a}}$$

$$E = KE + PE = \frac{e^2}{2a} - \frac{e^2}{a} = \frac{-e^2}{2a} = -\frac{me^4}{2\hbar^2} = -13.6\ eV$$

The $KE(r)$ term is positive everywhere – as it should be – since it is the kinetic energy of the electron. But, $K(r)$ is positive for $r < 2a$, zero for $r = 2a$ and negative for $r > 2a$. Note that it is the negative part of $K(r)$ that yields the overall negative electron energy. Both $KE(r)$ and $K(r)$ integrated over space yield the same total kinetic energy but, clearly, $K(r)$ does not represent the electron kinetic energy. In conclusion, the postulated "kinetic energy operator" does not yield the local kinetic energy of the particle. This wrong result does not affect calculations, since $K(r)$ yields the correct total kinetic energy and does not couple to other physical quantities.

An important insight that can be drawn from the expressions in (12) is that even though the potential is singular at the origin, the electron wavefunction, kinetic energy, and energy are finite and smooth functions. The wave properties of the electron preclude singularities.

## 5. CRITIQUE OF TRADITIONAL QUANTUM MECHANICS

The traditional quantum mechanics has been founded on local dynamical principles analogous to those of classical mechanics. However, while classical mechanics is rightly a local theory, since the motion of bodies change only at the points where forces act, quantum mechanics cannot be just a local theory because particles are described by spread out waves.

Erwin Schrodinger tried, in 1926, to find an "equation of motion" that would describe the particle waves proposed by Louis de Broglie. He derived the equation that bears his name, but on inappropriate grounds. As a result, his derivation became irrelevant and, since no one came up with a definite derivation, the Schrodinger equation became an axiom of quantum physics that gives successful results, but there is no adequate explanation why.

However, the simple and telling form of the Schrodinger equation indicates that the traditional approach is not founded on sound principles that would show the physical origin of this equation. This conclusion and the wrong result obtained by applying the "kinetic energy operator" call for a properly founded quantum theory.

# 6. FUNDAMENTAL PRINCIPLE OF QUANTUM MECHANICS

Physical theories are based on a few principles underlying all related phenomena. For example, classical mechanics is founded on Newton's second law of motion, evident every time a ball is kicked and discernible in the motions of all terrestrial and celestial bodies.

We can identify the principles of particle interactions by analyzing simple cases. For example, the interference pattern produced in the double slit experiment leads to the conclusion that it arises from waves spreading out of both slits simultaneously. These waves must be virtual since the photon (or electron) does not split in separate pieces; otherwise, we could catch half a photon (or a hundredth, if there are one hundred slits) – which does not happen. By moving the film plate closer or farther from the slits, the interference pattern contracts or expands, respectively, indicating that the photon waves have constant amplitudes. We also realize that the intensity of the total particle wave at a point on the film corresponds to the probability that the particle may be there. These features are the foundations of the quantum theory.

Before we proceed, we have to determine the exact form of the particle wave. Even though particle waves are usually depicted by single waving lines, this description is not correct because particles do not oscillate – we never detect less or more than a whole photon or electron. Therefore, we have to use some kind of wave that has constant intensity and, yet, its wave properties come alive when there are other waves of this particle. We achieve this by employing a double wave, whose parts $\psi_1$ and $\psi_2$ oscillate so that their superposition $\Psi$ has constant intensity: $\Psi^2 = \psi_1^2 + \psi_2^2$. Considering $\psi_1$ and $\psi_2$ along perpendicular directions, $\Psi$ is their vector sum, an arrow that rotates with period $T = h/E$ while its phase angle advances counterclockwise in the direction of the momentum completing a full rotation every $\lambda = h/p$ meters. In practice, we represent one wave with a real *cosine* function and the other with an imaginary *sine* function. The sum of these waves can be written in the compact exponential form shown in the integrand of the top equation in (13).

From our analysis, we determine that the fundamental principle of quantum mechanics is:

> Particles have multiple virtual motions and every state of motion (**p**,*E*) is accompanied by a wave that has constant amplitude $a(\mathbf{p},E)$. The wavefunction $\psi(\mathbf{r},t)$ is the superposition of all the virtual waves of the particle.

The intensity $|\psi(\mathbf{r},t)|^2$ of the wavefunction yields the probability that the particle may be at point (**r**,*t*). Likewise, $|a(\mathbf{p},E)|^2$ represents the probability of finding the particle in the (**p**,*E*) state.

The multiple states of a particle are virtual, but have real effects. For example, the potential energy of an electron is $PE(\mathbf{r}) = |\Psi_1(\mathbf{r})|^2 V(\mathbf{r})$, as if a $|\Psi_1(\mathbf{r})|^2$ portion of the electron is at **r**. Upon measurement, only one of the multiple virtual states materializes. By repeated measurements, we obtain $|a(\mathbf{p},E)|^2$, being the portion of the times we find the particle in the (**p**,*E*) state.

According to its fundamental principle, quantum mechanics is a local and global theory. In particular, the space-time variation of the wavefunction results from the interference of constant amplitude waves that arise from the interactions of the particle.

Since the wavefunction is a superposition of particle waves, and each wave carries a specific momentum (**p**) and energy (*E*), the wavefunction also yields the particle momentum and energy, as well as its kinetic energy, angular momentum, etc. at each point. For example, the particle momentum **p**(**r**,*t*) is the sum of the products of the momentum **p** of each wave times the wave amplitude at (**r**,*t*). Likewise, the expressions shown below give the local or total values of energy, kinetic energy, angular momentum, etc.

$$\psi(\mathbf{r},t) = \frac{1}{(2\pi\hbar)^{4/2}} \int a(\mathbf{p},E) e^{\frac{i}{\hbar}(\mathbf{p}\cdot\mathbf{r}-Et)} d^3\mathbf{p}\, dE$$

$$\mathbf{p}(\mathbf{r},t) = \frac{1}{(2\pi\hbar)^{4/2}} \int \mathbf{p}\, a(\mathbf{p},E) e^{\frac{i}{\hbar}(\mathbf{p}\cdot\mathbf{r}-Et)} d^3\mathbf{p}\, dE = -i\hbar \nabla \psi(\mathbf{r},t)$$

$$\mathbf{p} = \frac{1}{(2\pi\hbar)^{4/2}} \int \mathbf{p}|a(\mathbf{p},E)|^2 d^3\mathbf{p}\, dE = \frac{1}{(2\pi\hbar)^{4/2}} \int \psi^*(-i\hbar\nabla\psi) d^3\mathbf{r}\, dt \qquad (13)$$

$$KE(\mathbf{r},t) = \frac{p^2(\mathbf{r},t)}{2m} = \frac{1}{2m}\left(+i\hbar\nabla\psi^*\right)\cdot\left(-i\hbar\nabla\psi\right) = \frac{+\hbar^2}{2m}\nabla\psi^*\cdot\nabla\psi$$

$$E(\mathbf{r},t) = \frac{1}{(2\pi\hbar)^{4/2}} \int E\, a(\mathbf{p},E) e^{\frac{i}{\hbar}(\mathbf{p}\cdot\mathbf{r}-Et)} d^3\mathbf{p}\, dE = i\hbar\frac{\partial}{\partial t}\psi(\mathbf{r},t)$$

$$L_z(\mathbf{r},t) = x p_y(\mathbf{r},t) - y p_x(\mathbf{r},t), \quad etc.$$

Quantum theory specifies the values of the particle momentum, energy, etc. at each space-time point without clashing with the uncertainty principle. To see this, consider a wavefunction consisting of three waves $\psi_1, \psi_2, \psi_3$ with momenta $p_1, p_2, p_3$. A measurement will yield any one of these momentum values, but we would not know at what point the particle was; this is consistent with the fact that $a(\mathbf{p},E)$ does not have any space-time dependence. Nevertheless, the value of the particle momentum at point $x$ is $p(x) = p_1\psi_1(x) + p_2\psi_2(x) + p_3\psi_3(x)$. This theoretical description gives correct results when the interaction depends on $p(x)$.

## 7. PARTICLE INTERACTIONS

We can describe and explain many particle interactions by figuring out what are the alternative virtual motions (or paths) of the particle and perceive how their waves add up.

In scattering processes involving simple boundaries – e.g. double slits, reflecting and refracting surfaces (rows of atoms), potential barriers, even potentials – it is easy to visualize the possible alternative motions, add their waves and obtain – qualitatively and quantitatively – the probabilities of the various outcomes.

Bound particles have zero total momentum, therefore, they are in a limbo superposition of virtual motions that have opposite momenta. For a particle in a box, the momenta (wavelengths) of the virtual motions are determined by the size of the box. The electron in a Hydrogen atom can have any momentum (since there are no boundaries) but the amplitudes of its virtual waves vary. We can measure these amplitudes by scattering photons off Hydrogen atoms. From such "photoelectric" processes we get the momenta of the electron virtual states and measure how frequently they come up. Thus, we find the amplitudes of the virtual electron waves to be $|a(\mathbf{p})|^2 \propto \left(1 + b^2 p^2\right)^{-4}$, where $b^2 = \hbar^2(m^2 e^4)^{-1}$. Inserting $a(\mathbf{p})$ in the top equation of (13), we obtain the electron wavefunction $\Psi_1$ shown in (12). We can also calculate the wavefunction theoretically. We will see that the total energy relation requires that the electron wavefunction $\Psi_1$ must satisfy at all points the Schrodinger equation, which we solve to get $\Psi_1$.

Ultimately, particles interact by exchanging intermediate particles and infinite (but few significant) processes contribute to each outcome. According to the overarching principle of quantum mechanics, we obtain the probability of an outcome by adding together the waves of all contributing processes.

## 8. DERIVATION OF THE SCHRODINGER, ETC. EQUATIONS

Consider a particle interacting with an appropriately continuous potential $V(x)$. The particle energy, kinetic energy and potential energy at point $x$ are:

$$E(x) = E|\psi(x)|^2 \qquad KE(x) = \frac{+\hbar^2}{2m}\frac{\partial \psi^*}{\partial x}\frac{\partial \psi}{\partial x} \qquad PE(x) = V(x)|\psi(x)|^2 \qquad (14)$$

The non-relativistic (local) energy equation $E(x)=KE(x)+PE(x)$ is not satisfied. For example, we can see this explicitly from the expressions in (12).

The total energy equation is

$$E = KE + PE \quad \rightarrow \quad \int E|\psi|^2 dx = \int \frac{+\hbar^2}{2m}\frac{\partial \psi^*}{\partial x}\frac{\partial \psi}{\partial x} dx + \int V(x)|\psi|^2 dx \qquad (15)$$

Using (8), we can put the above equation in the following forms:

$$\int \psi^* E\psi\, dx = \int \frac{-\hbar^2}{2m}\psi^*\frac{\partial \psi}{\partial x}dx + \int \psi^* V(x)\psi\, dx$$

$$\int \psi^* \left[ E\psi + \frac{\hbar^2}{2m}\frac{\partial^2 \psi}{\partial x^2} - V(x)\psi \right] dx = 0 \qquad (16)$$

Next, we show that the integrand in (16) is zero at all points because the wavefunction $\psi(x)$ is the sum of virtual waves that have constant amplitudes. [Note that since $\psi(x)$ is a sum, we cannot factor out the amplitudes of its waves.]

If the integrand at point $x_1$ were equal to the non-zero value $N_1 = \psi^* n_1 \psi$, we would have:

$$\psi^*\left[ E\psi + \frac{\hbar^2}{2m}\frac{\partial^2\psi}{\partial x^2} - V(x_1)\psi \right] = \psi^* n_1 \psi \quad \rightarrow \quad E\psi + \frac{\hbar^2}{2m}\frac{\partial^2\psi}{\partial x^2} - [V(x_1)+n_1]\psi = 0 \qquad (17)$$

The above equation requires that the constant amplitudes of the virtual particle waves depend on the arbitrary $n_1$ parameter. Likewise, if the integrand were not zero at $x_2$, $x_3$, ..., the amplitudes would also depend on the arbitrary parameters $n_2$, $n_3$, ... which, of course, is unacceptable. Therefore, the integrand in (16) vanishes everywhere and dividing it by $\psi^*$ we get the Schrodinger equation:

$$E\psi = \frac{-\hbar^2}{2m}\frac{\partial^2\psi}{\partial x^2} + V(x)\psi \qquad (18)$$

In conclusion, the Schrodinger equation has been derived to be the (local) condition the wavefunction must satisfy at each point in order to fulfill the total (global) energy equation.

In an analogous fashion, we can derive the three dimensional, time-dependent Schrodinger equation and also the other wavefunction equations from the respective total energy equations. In particular, we derive the Dirac equation from the linearized relativistic energy equation that describes a spin-1/2 particle interacting with a four-potential $A_\mu(\mathbf{r},t)$. Since the squares of energy and momentum are not involved, the Dirac equation corresponds to a local energy equation and can be put, validly, in the "time-evolution and Hamiltonian" form shown in (2).

## 9. DISCUSSION

Subatomic particles are associated with spread out waves. Consequently, while classical mechanics is a local theory, quantum mechanics is a local and global one. We clarify the difference between these two theoretical foundations below.

A macroscopic body is at a single location at each moment, and changes state of motion only at the points where a force acts on it. The trajectories of bodies are described by classical mechanics, which is a local theory, meaning that cause and effect happen at the same space-time point. For example, a rolling ball bounces only at the point it hits a bump, neither before nor after. In contradistinction, the insertion of a little bump inside a box containing an electron alters the electron wavefunction everywhere and continuously. In general, every detail of the interaction has global effects. In the non-relativistic domain, the effects of every interaction feature manifest everywhere automatically, while in the relativistic quantum interaction theory all interaction aspects contribute to the probabilities of the outcomes.

The traditional quantum mechanics was developed as a local theory and, as a result, it does not provide tangible descriptions of particle phenomena; instead, it is based on (local) postulated equations. A central postulate is the Schrodinger equation, which is assumed to be an energy-equals-kinetic-plus-potential-energies equation. We have shown that this is a wrong assumption because the term that contains the second space derivative does not represent the particle kinetic energy. This error led to distorted understandings of the Schrodinger equation and prompted inept heuristic derivations.

Richard Feynman declared that "you cannot derive the Schrodinger equation no matter how hard you may try" [4]. However, the impossibility of deriving the Schrodinger equation has nothing to do with its difficulty – after all, it has a simple and telling form – but with the theoretical framework. The fact that the Schrodinger equation cannot be derived, and the error arising from applying the "kinetic energy operator" indicate that the traditional space-time approach (including the path-integral formulation) is not based on principles that reflect properly the character of particle interactions. Responding to Feynman's admonition, we may point out that a theory that cannot derive simply and directly the Schrodinger equation does not have proper physical foundations.

Taking a fresh look at particle interactions, we perceive that particles have multiple simultaneous virtual motions and that each motion is associated with a wave that has constant amplitude. Accordingly, the space-time variations of the wavefunction result from the superposition of all the waves of the particle. In addition, the particle momentum, energy, potential energy, etc. have specific local and total values, which may not be simply related due to their wave dependencies. In particular, the total energy is equal to the total kinetic plus potential energies but, locally, this relation may not hold. In the non-relativistic approximation, the wavefunction satisfies at each point a different energy condition that corresponds to the Schrodinger equation.

In conclusion, the theory of quantum mechanics is local and global and conceptually simple: we just add the waves of all possible particle motions. Based on this fundamental principle, we can describe many particle interactions in tangible terms and provide a firm, non-mathematical presentation of quantum mechanics. Furthermore, the mathematical foundation of quantum physics becomes clear and justified, and the wavefunction equations (Schrodinger, Pauli, Dirac, etc.) are derived to be the conditions the wavefunction must satisfy at each point in order to fulfill the respective total energy equation.

10## REFERENCES

Scores of quantum mechanics books could be listed below. Instead, we chose four prominent texts that expound the typical attitudes and treatments of the traditional quantum theory.

[1] Griffiths, D., *Introduction to Quantum Mechanics*, 2nd ed., Prentice-Hall, New Jersey, 2004.
[2] Greiner, W., *Quantum Mechanics an Introduction*, Springer, New York, 1994.
[3] Weinberg, S., *Lectures on Quantum Mechanics*, Cambridge University Press, Cambridge, 2012.
[4] Feynman, R., Leighton, R., Sands, M., *The Feynman Lectures of Physics*, Addison-Wesley, Boston, 1964.